# Individual Barkhausen pulses of ferroelastic nanodomains

Reinis Ignatans, Dragan Damjanovic, and Vasiliki Tileli*

Institute of Materials, École polytechnique fédérale de Lausanne, Station 12, 1015 Lausanne, Switzerland.

*Correspondence to: vasiliki.tileli@epfl.ch

Ferroelectric materials, upon electric field biasing, display polarization discontinuities known as Barkhausen jumps, a subclass of a more general phenomenon known as crackling noise. Herein, we follow at the nanoscale the motion of 90º needle domains induced by an electric field applied in the polarization direction of the prototypical ferroelectric $BaTiO_3$, inside a transmission electron microscope. The nature of motion and periodicity of Barkhausen pulses leads to real-time visualization of distinctive interaction mechanisms of the domains with each other but without coming into contact, a mechanism that has not been observed before, or/and with the lattice where the domain walls appear to be moving through the dielectric medium relatively freely, experiencing weak Peierls-like potentials. Control over the kinetics of ferroelastic domain wall motion can lead to novel nanoelectronic devices pertinent to computing and storage applications.

Polarization switching in ferroelectric crystals is akin to first order phase transition induced by an applied electric field (*1*). The re-orientation of the domains during switching is a response to assume the equilibrium conditions that minimise the free energies of the domain and domain wall formation and growth. The motion towards the equilibrium state is hindered by potential barriers that lead to non-monotonous polarization change discontinuities referred to as Barkhausen effect (*1, 2, 3, 4, 5, 6*). In ceramic (*5*) and single crystal $BaTiO_3$ (*3, 7*) the origins of this effect is placed on nucleation of spike-like domains, pinning of domain walls due to defects in the lattice (*2, 6*), domain coalescence (*8*), and transition of the needle-like domains to parallelepiped domains when they fully cross the crystal in the direction along which the field is applied (*7*). Most electrical studies of Barkhausen jumps attempt to resolve single events from macroscopic properties of ferroelectrics (such as switching current or charge) while in optical observations the temporary position and motion of domain walls is probed. The various methods may reveal different origins of the pulses. This is associated with varying measurement timescales, different boundary conditions of the ferroelectric and physical dimensions of the material. At nanoscale dimensions traditional electrical measurements become exponentially more difficult due to miniscule charges associated with individual switching while optical measurements are diffraction limited. Kinetically-controlled mechanisms of nanodomain switching and the way their interaction affects their motion when the dimensions become small have not been reported yet.

In the following, we stabilize the domain structure of a thin single crystal $BaTiO_3$ to fully contain 90º ferroelastic nanodomains and we track their growth by applying a well-oriented external electric field. We locally probe the Barkhausen jumps for different potential barriers and at different electric field frequencies.

To achieve this, we prepared lamellae specimen along the pseudocubic $[001]_{PC}$ orientation from a single crystal $BaTiO_3$ and placed it on a microelectromechanical system (MEMS) chip patterned with six electrodes; four used for temperature control and two for electric field biasing. The geometry of the device (Fig. S1 and Materials and Methods in supplementary materials) permits application of homogeneous electrical field across the probed sample area (*9*).

First, the system was heated from room temperature over the ferroelectric-to-paraelectric transition temperature ($T_C$) before cooling at 130 ºC for the biasing experiments (the full heating profile and image series is shown in Supplementary Movie S1). Sequential bright field (BF) scanning transmission microscopy (STEM) images of the temperature-induced process are depicted in Fig. 1, A to D. The $T_C$ was determined by the appearance and disappearance of the domains and it was measured to be 148 ºC on heating and 137 ºC on cooling. A systematic overestimation of the measured temperature of up to 18 °C (usually $T_C \sim$ 131 ºC from macroscopic measurements (*10*)) is attributed to error in the local temperature control of these devices. It is noted that heating above Tc was required to reduce the strain, which inevitably occurs due to the FIB preparation procedure (*11*). At room temperature, the specimen is dominated by unusually stable 180º walls (*12*), Fig. 1A, whereas during heating, a transition to the 90º domain structure at T ~ 50 ºC (*13*) is induced, Fig.1B. We note that this behaviour is not associated with the strain from clamping the lamella from both sides on the MEMS chips, as the same process is seen in a free-standing lamella as shown in Fig. S2.

Fig. 1D depicts the domain structure under the Curie temperature which consists of periodic 90° ferroelastic needle domain groups. To accurately characterize the ferroelastic domain structure at 130 °C, differential phase contrast (DPC) imaging (*14*) was employed. The DPC signal strength and direction is directly proportional to the magnitude and direction of the local polarization (*14*). Fig.1E represents the local profile of polarization directions encoded by arrows from the square region shown in Fig.1D. The size of the arrows is proportional to the magnitude of the polarization, however, absolute values are difficult to report since minor alignment issues greatly affect the DPC results. Nevertheless, a closer look, Fig.1F, illustrates that the 90° domain walls in tetragonal BaTiO$_3$ align with the $[110]_{PC}$ direction (*1*), as expected by crystallographic symmetry laws. The average domain width, *w*, is measured to be about 33 nm. The domain wall is less than 7 nm thick and it is associated with dark contrast (i.e. high angle scattering) in BF-STEM images. The appearance of the periodic domain structure in thin films and slabs is associated with the thickness of the slab, $w^2 \propto d$, where *w* is domain width and *d* is slab's thickness (*15*). The same proportionality stays true for the ferroelastic domain structure (*16*, *17*, *18*, *19*).

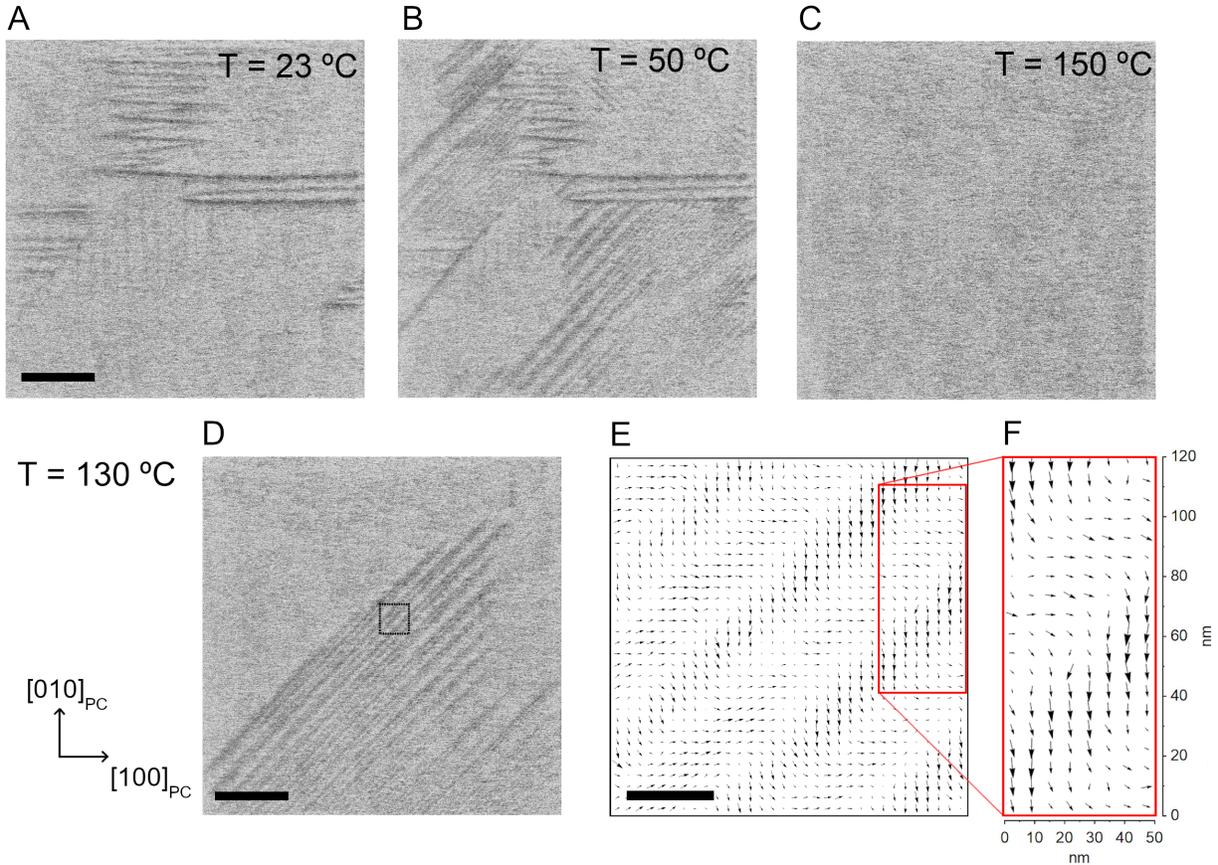

**Fig. 1. Domain evolution upon heating and subsequent cooling.** (**A**) Domain structure of BaTiO$_3$ at room temperature showing 180° domain walls. (**B**) Domain structure of BaTiO$_3$ during transition between 180° and 90° domains at 50 °C. (**C**) Paraelectric (cubic) BaTiO$_3$ at 150 °C. (**D**) Domain structure at 130 °C showing periodic ferroelastic 90° needle domains (scale bar is 500 nm). (**E**) DPC image of the black dotted square area in **D** (scale bar is 50 nm). The arrows represent the polarization direction. (**F**) Close-up of polarization vectors of the ferroelastic nanodomains.

At 130 ºC (nominal temperature), we applied a triangular waveform voltage (see Methods and Fig. S3) and we followed the responses of several 90º, ferroelastic, a-type needle-like domains. Fig. 2A shows the domain structure at relevant points for maximum applied bias of 3.5 Volts and the full sequence can be seen in Supplementary Movie S2. Two distinct cases are plotted as domain length vs. applied potential in Fig. 2, B and C. These highly resemble traditional polarization-electric field (P-E) loops, confirming that the observed effect is electric field induced whereas macroscopic strain-induced effects have been minimized. While conventionally measured P-E loops show macroscopically averaged behavior, here we are able to follow the response locally on a single domain level, where polarization is proportional to the length of the domain. Interestingly, we have observed two distinct manifestations of the needle domain response on the electric field. When the vertex of a needle domain is located further away from the crossing domain walls (Fig 2B), the measured domain length vs. applied potential loop is smeared and the process is lattice defect mediated (*20*, *21*). On the other hand, when the needle domain's vertex is located close to the crossing domain walls (Fig. 2B) the resulting shape of the domain length vs. applied potential loop is sharp and square-like. Therefore, this domain-domain interaction mediated process seems to involve strong local strain and depolarizing fields that lead to the domain vertex tracing a path with larger hysteresis as it goes back and forth, unlike the defect-mediated process where the effect of the lattice and possible mobility of defects lead to a less hysteretic behavior.

Closer investigation of the loop in Fig. 2C shows clear step-like features around, for example, 1 and 2 V, whereas for needle domains whose vertices lie further away from the perpendicular needle domain walls (Fig. 2B), step-like features are less pronounced. Thus, needle domains whose vertices are close to the perpendicular crossing domain walls, experience more pronounced non-monotonous movement during biasing (*5*). These Barkhausen jumps occur at the moments when the crossing needle domain has contracted. It is noted that crossing needle domains do not interact directly with perpendicular domains (i.e. they are not in direct contact with them), but rather interaction of needle domains is mediated through a large parent domain (the background grey domain in the images of Fig. 2A). The equilibrium position of the needle domains between Barkhausen jumps is associated to both electric and mechanical compatibility (*22*, *23*, *24*). Two possibilities can be distinguished, the first one, where the needle domain's vertex terminates next to the crossing needle domain's body, and the second, in which equilibrium is achieved if two or more strongly charged needle domain vertices come in close contact (*22*). The observed Barkhausen jumps occur in between those equilibria. We note that experiments for an increasing maximum range of the applied bias were additionally performed and similar domain behaviour was observed (Supplementary Movies S3 to S8).

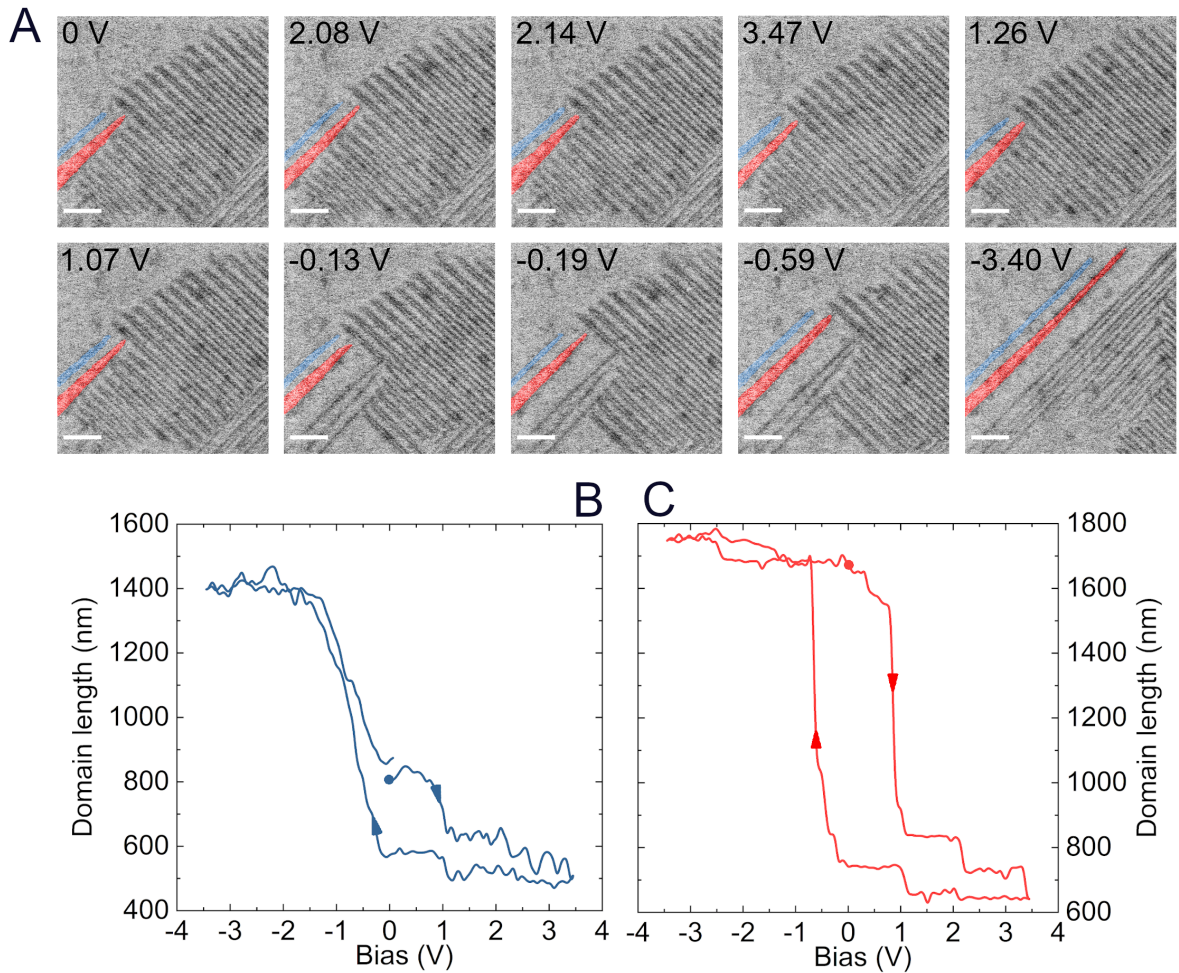

**Fig. 2. Ferroelastic domain motion with respect to applied electric field at 130 °C.** (**A**) 90º domain structure evolution during biasing. The measured domains have red and blue shading. Scale bar is 300 nm. (**B**) Domain length as a function of applied potential plots of a domain with weak interaction with crossing domain walls (blue false color in **A**). (**C**) Similar loop of a domain that interacts strongly with crossing domain walls (red false color in **A**). The starting point of the loops is marked with a dot and the loop rotation direction is marked with arrows.

To study the time dependence of the needle domain response, similar experiments were carried out with a less steep voltage ramp (0.0034 V/s) corresponding to ultra-low frequency of 0.24 mHz. The hysteresis loop can be seen in Fig. S4 and the full sequence is shown in Supplementary Movie S9. The overall domain length vs. applied voltage loop is steep and square-like and it is similar to the one in Fig. 2B that represents strong domain/domain interaction. Again, Barkhausen jump events associated with interacting needle domain equilibria positions were recorded.

To compare the response of the processes leading to characteristically different domain wall motion, their length is plotted as a function of time, Fig. 3A. Both time axes in Fig. 3A have been rescaled to match the start and the end of the bias cycle in a single graph. The polarization of the domain during the low frequency measurement is parallel with the applied electric field's direction, which results in domain growth with positive voltage and shrinkage with negative applied voltage (i.e. response is opposite to the rapid experiment where the domains nucleated with opposite polarization). Clear differences in the Fig. 3A can be seen between lattice defect mediated needle domain growth (blue squares) and the domains which experience strong domain-domain interaction (red and green squares). The latter ones exhibit distinctive step-like Barkhausen jumps (marked in the Fig. 3 with green and red arrows), whereas domain growth via lattice mediated defect mechanism is considerably smoother. The key difference observed during the low frequency measurement is the appearance of the domain structure relaxation events with a time constant of roughly three to five minutes. For example, one such event can be seen in Fig. 3A between 2500 and 2900 s. The measured domain first decreases upon applied negative voltage, then it slowly increases in length (marked in Fig. 3A with grey horizontal line) and eventually exhibits a Barkhausen jump around 2900 s, before further shrinkage of the domain. After this jump, the domain again slightly increases in length, showing similar behavior as before. Interestingly, although at around 3000 – 3300 s negative potential is slowly returned to zero volts, the domain annihilates completely (marked with the dashed green arrow in the Fig. 3A). In short, the domain experiences larger applied electric field and stays intact, but on the decreasing field it slowly decreases in size and disappears, which again shows a relaxation event with a time constant of several minutes. The observed sluggish relaxation processes can be attributed to dielectric viscosity (*2*). According to observations from electrical measurements, the domain structure can relax up to several hours after poling. We may therefore witness here an individual event responsible for ageing and creep in ferroelectric materials (*25*, *26*, *27*).

When the lower frequency domain motion is compared to higher frequency ones, we notice a time-delay of the switching process when the potential is brought back to zero. During ultra-low frequency measurements, the domain remains at the same position for a significant time before it eventually switches.

By numerically differentiating the domain length vs. time data (Fig. 3A), the speed of the needle domains can be determined. Figure 3B shows the velocity as a function of applied voltage for the three distinct domains. Essentially, the speed is proportional to the switching current (i.e. $i \sim \frac{\Delta P}{\Delta t}$) and here we demonstrate the local speed in the motion of a single domain. The two peaks of the domain/domain interaction case (red lines) represent voltage where the applied electric field is close to the coercive field. By comparing the lattice defect mediated growth (blue line) with the domain-domain interaction one, it is determined that the domain tip velocity for the defect-mediated domain is more smeared out (i.e. the vertex is almost always moving). In contrast, the velocity of domain-domain interaction tip is accompanied with spikes and, at most times, the velocity is close to zero. This indicates that the potential well is deeper for domain-domain associated pinning as compared to pinning on the defects within the lattice (*28*, *29*). Additionally, in the case of the low frequency measurement (green line), the velocity spikes can be seen in the deep negative potential part representing Barkhausen jumps due to domain-domain interaction and slow relaxation processes simultaneously.

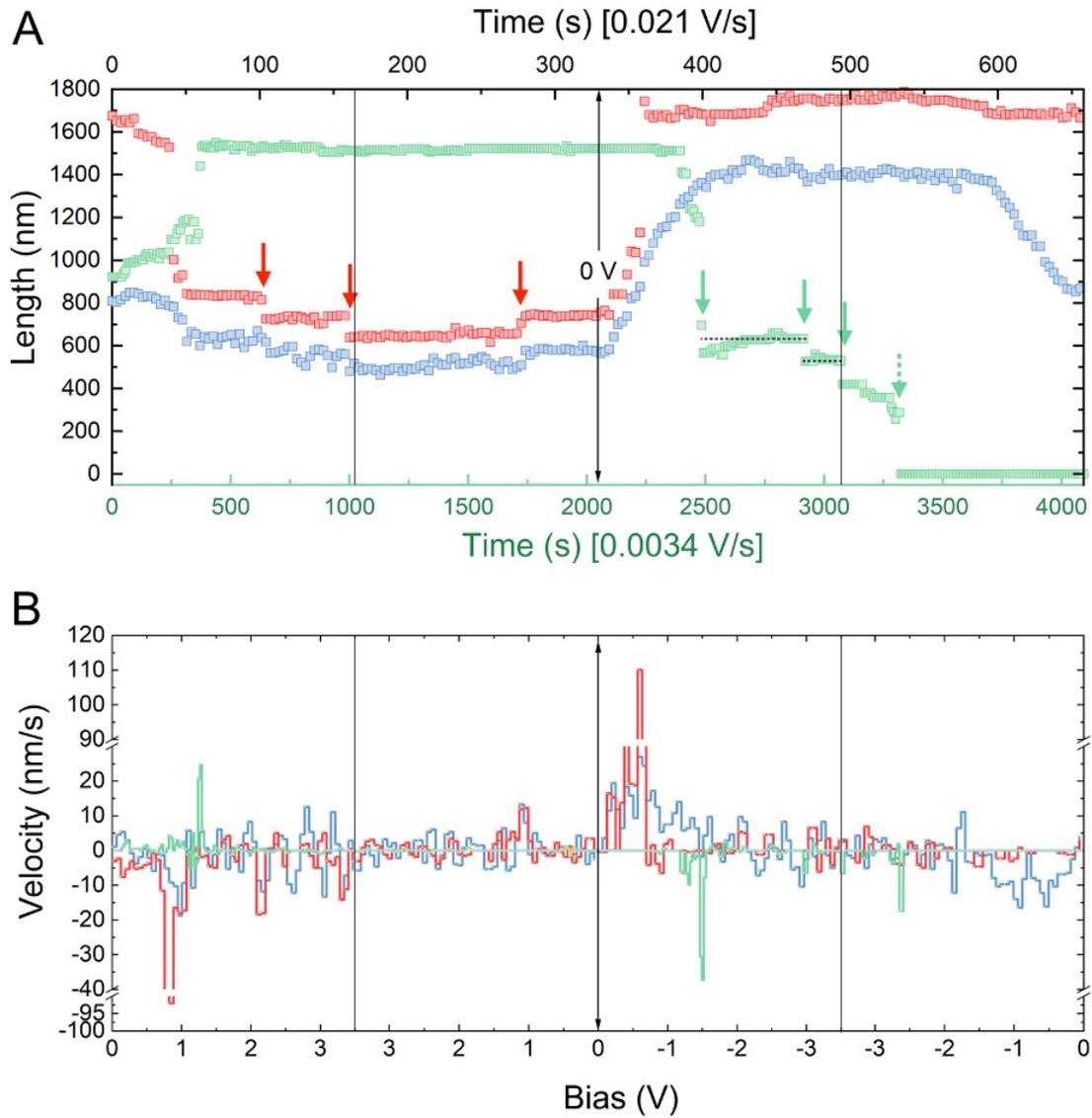

**Fig. 3. Domain evolution and speed at different frequencies.** (**A**) Time domain of the domain length measurements with respect to applied voltage. The measurements of the two domains (red

and blue of Fig. 2A) correspond to the top black x axis (performed at 0.021 V/s) whereas the green squares and bottom x axis correspond to the slow domain response measured in Fig. S4 (performed at 0.0034 V/s). Red and green vertical arrows represent Barkhausen events. The final, dashed, green arrow represents the point of annihilation of the domain. Grey dotted horizontal lines represent domain length after relaxation. (**B**) Domain tip velocity with respect to applied bias voltage, color coding is the same as in **A**. Time and bias x axes on the two plots are scaled in such way that they correspond to each other on the basis of the field waveform.

Overall, the observed evolution and motion of ferroelectric single needle domains under application of electric field in the polar direction is characteristic of a forward domain growth process (*30*). However, we have previously shown that such movements follow Rayleigh-like square-law behaviour (*9*) and, therefore, pinning mechanisms typically studied for lateral domain wall movements are also applicable in this high field regime. To further assess the mechanisms of pinning of 90º needle nanodomain wall motion induced by external electric fields, we discuss their dynamic behaviour on the basis of the schematics in Fig. 4. It was recently shown that the equilibrium positions of 90º needle domains in $BaTiO_3$ are adopted due to the redistribution of strain and electric displacements fields on their tips (*22*). Our results show that, in the case of a herringbone domain pattern, this effect causes separation of crossing domains that terminate at a certain distance from the perpendicular crossing domain without ever coming into contact (Fig. 4A). Moreover, the depolarizing and strain fields caused by the abrupt changes in the spontaneous polarization at the needle tip form a large potential barrier that manifests as well-defined Barkhausen jumps. The positions of these jumps are associated with the annihilation of the periodic, perpendicular domains, and the domain-domain induced pinning motion leads to hysteretic behaviour. A different mechanism is encountered when parallel needle domains are free to move within the lattice structure (Fig. 4B). In this case, the Barkhausen pulses are rare events, and are most likely dominated by Peierls potential barriers due to lattice potential and point defects in the non-perfect crystal (*28*, *29*). The associated shallow potential barrier does not affect perpendicular domain-domain motion but is dominating the jump frequency during the motion of single domains across the lattice. Therefore, at the moderate electric field and frequencies used herein, classical Barkhausen jumps mostly originate from domain-domain interactions and much less from interaction of domains with point defects.

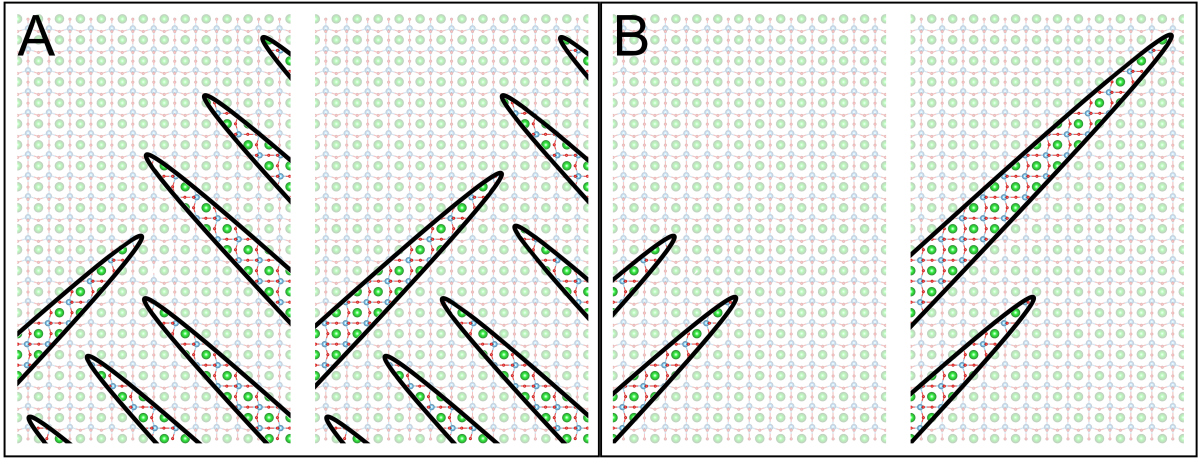

**Fig. 4. Mechanisms of 90º needle nanodomain motion driven by external electric field in the polar direction.** (**A**) Needle domain configuration in a metastable electromechanical equilibrium before a Barkhausen event and new equilibrium after the application of external electric field in the polar direction. The distance of the jump is directly associated with the periodic domain structure of the needle domains. (**B**) Configuration of two needle domains located far from perpendicular, crossing domain walls. Their movement and pinning are associated with the shallow Peierls potential of the atomic lattice containing few defects and, therefore, the Barkhausen jumps do not occur at distinct characteristic distances due to their relatively free movements. The electric field direction is from right to left in both cases.

In conclusion, we have demonstrated conjunct ferroelastic needle domain response on the applied electric bias field leading to strong pinning and consequential Barkhausen jumps in between domain structure metastable equilibria. The mechanisms of domain-domain interactions among non-contacting domains remain largely unexplored. Our experimental study shows that major domain pinning and restriction to movement in a thin single crystal BaTiO$_3$ ferroelectric is predominantly associated with mutual domains interaction and less with Peierls potential pinning arising from the lattice in an undoped crystal. The shape of the measured domain length-electric bias loops hints that domain-domain interaction may be dominant in the expression of the materials properties macroscopically. Individual local relaxation events leading to ageing and creep have also been observed. Such insights gained from local measurements performed in a thin ferroelectric slab, can be relevant for modern technologies related to polarization switching.

**ACKNOWLEDGEMENTS**

The work was supported by the Swiss National Science Foundation (SNSF) under award no. 200021_175711. D. D. acknowledges support by the ONR Global (award No. N62909-18-1-2078).

**Supplementary Materials:**

Materials and Methods

Figures S1-S4

Movies S1-S9

# Supplementary Materials for

## Individual Barkhausen pulses of ferroelastic nanodomains

Reinis Ignatans, Dragan Damjanovic, and Vasiliki Tileli*

Correspondence to: vasiliki.tileli@epfl.ch

**This PDF file includes:**

    Materials and Methods
    Figs. S1 to S4
    Captions for Movies S1 to S9

**Other Supplementary Materials for this manuscript include the following:**

    Movies S1 to S9

**Materials and Methods**

Sample preparation

Single crystal BaTiO$_3$ (MTI Corporation) was focused ion beam prepared and it was transferred to a 4 heating and 2 biasing electrode microelectromechanical (MEMS) chip provided by DENSsolutions. The Pt contacts were ion beam deposited and the lamella was thinned to a final thickness of 268 nm. A detailed description on the device fabrication can be found in Ref. [9]. The average temperature on the heating element of the MEMS chip is calculated from its electrical resistance and DENSsolutions software was used to control the temperature. Electrical bias to the sample was applied using a Keithley SMU-2450 sourcemeter.

Transmission electron microscopy

Imaging was performed on a double spherical aberration (C$_s$) corrected ThermoFischer Scientific Themis 60-300 operated at 300 kV in STEM mode using 100 pA beam current, 70 µm C2 aperture, and a beam convergence angle of 28 mrad. Serial imaging during standard biasing experiments was done in bright field (BF) mode with a collection angle of 79 mrad. The pixel size was 3.2 nm and 2.98 s was the frame time. For the low frequency biasing experiment, the pixel size was 1.1 nm and the frame time was 8.22 s. DPC imaging was performed using a four segment detector and the imaging conditions were 200 pm pixel size and 34.3 s frame time. The camera length was 91.1 mm. Beam centering and gain/offset equilibration of the segments were done in the vacuum region prior to the experiment.

Image processing

Images of the standard biasing experiment were cropped to 764x764 pixel size. Images of the low frequency field measurement were not cropped and remained at the 2048x2048 pixel size. To enhance domain wall contrast in the BF images, postprocessing using ImageJ software was performed. It involved Fourier filtering of the horizontal scan noise of the images, subtraction of background (50 pixel rolling ball radius with sliding paraboloid enabled) followed by contrast and brightness adjustment.

DPC experimental signal processing involved the same 1024x1024 pixel region from the original set of 4 images (corresponding to the segments A, B, C, and D). The cropped region was further downsized with ImageJ to 256x256 pixel size essentially averaging out the noise. Further data processing was undertaken using a custom script written in Mathematica v.11.2. A 256x256 array was created, where each element represented a vector in the xy plane corresponding to the beam deflection. Beam deflection along the x axis is proportional to the differential signal D-B and correspondingly along the y axis it is proportional to C-A. Each element was corrected for the detector rotation with respect to the scan direction (+17°). Using this array, the vector field plots were created (Fig. 1e-f). The vector field plot consists of 32x32 arrows, which are proportional to the direction and magnitude of the polarization.

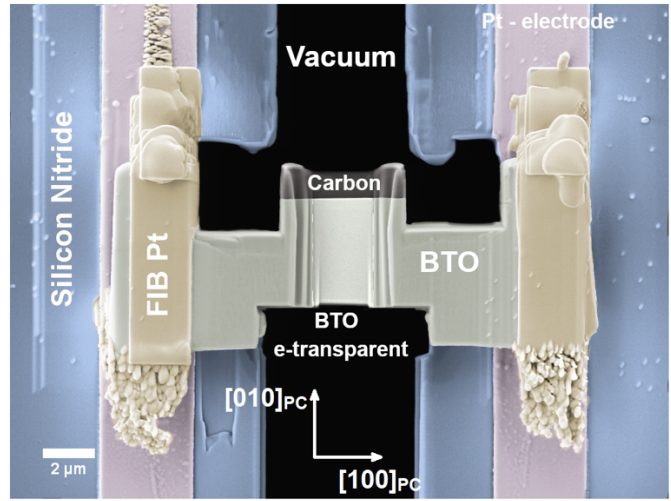

**Fig. S1. Sample geometry for the *in situ* experiments**. False coloured SEM image of the device used in the biasing experiments. Blue parts correspond to $SiN_x$ support membrane, pink vertical lines are Pt electrodes of the MEMS device, yellow shaded regions are FIB deposited Pt contact layers, light grey parts are BTO single crystal lamella (with the brightest part corresponding to the electron transparent region), dark grey represents the protective carbon layer and black corresponds to the FIB cut hole in the device.

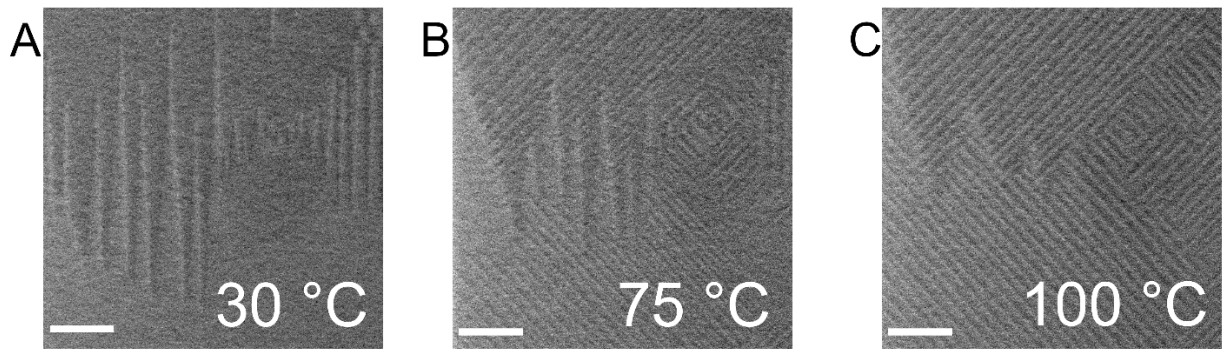

**Fig. S2.** *In situ* **heating experiment of free-standing lamella.** (**A**) Domain structure at 30 °C exhibiting 180° zig-zag domain walls. (**B**) At 75 °C, the sample is mostly dominated by 90° domain walls, nevertheless some character from zig-zag domain walls is still seen. (**C**) Domain structure at 100 °C where the specimen is dominated by 90° domain walls. The appearance of zig-zag wall vertices seems to be associated with crossing 90° domain walls. Scale bar is 250 nm. The experiment was performed on ThermoFisher Talos F200S microscope operated at 200 kV in STEM mode and imaged with the high angle annular dark field (HAADF) detector. The sample was mounted on a standard grid and a Gatan single tilt heating holder was used for heating the sample.

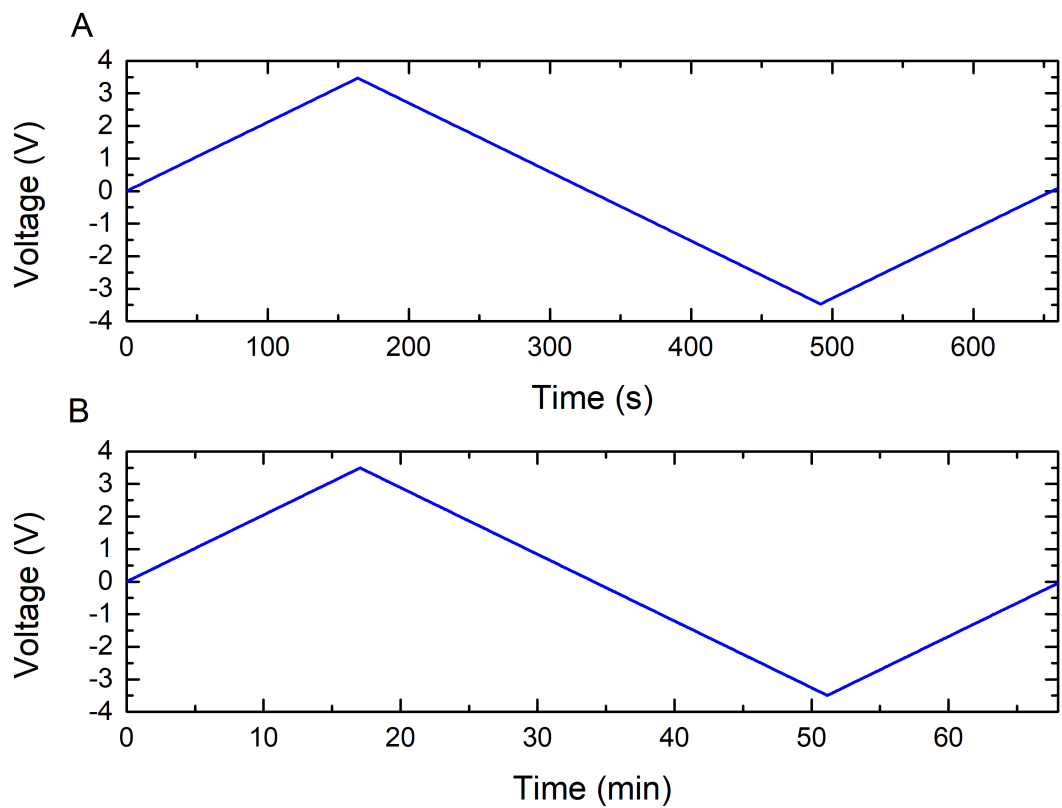

**Fig. S3. Triangular voltage profiles as a function of time.** (**A**) profile used for ferroelastic domain biasing of the domains in Fig. 2 and (**B**) profile used for the low frequency experiment.

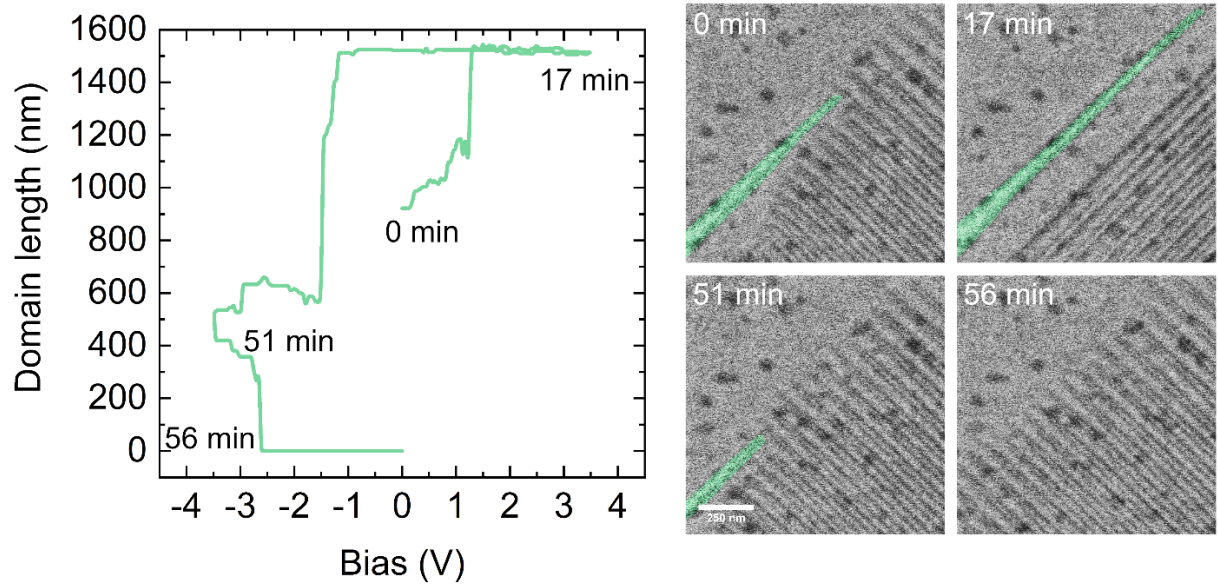

**Fig. S4. Low frequency domain motion.** Domain length as a function of applied bias at 130 °C, and image sequence of the domain at 0, 17, 51, and 56 min. The orientation of the loop is opposite as in Fig. 2 due to the measured domain (green false color) nucleating with antiparallel polarization compared to the ones in Fig. 2. The dark spots in the BF-STEM images are surface contamination, which grew upon lengthy *in situ* biasing and heating experiments. These spots seem to be located at the surface inactive layers and hence do not affect the domain motion.

**Captions of Movies S1-S9**

**Movie S1.**
Pre-heating of the BaTiO$_3$ lamella and cool-down to 130 °C before biasing experiment.

**Movie S2.**
Domain length – electrical bias loops and image series upon biasing with potential range of +/- 3.5 V.

**Movie S3.**
Domain structure response during biasing with potential range of +/- 0.5 V.

**Movie S4.**
Domain structure response during biasing with potential range of +/- 1.0 V.

**Movie S5.**
Domain structure response during biasing with potential range of +/- 1.5 V.

**Movie S6.**
Domain structure response during biasing with potential range of +/- 2.0 V.

**Movie S7.**
Domain structure response during biasing with potential range of +/- 2.5 V.

**Movie S8.**
Domain structure response during biasing with potential range of +/- 3.0 V.

**Movie S9.**
Low frequency domain response measurement with potential range of +/- 3.5 V.